\documentclass[twocolumn,notitlepage,superscriptaddress]{revtex4-1}

\usepackage{dcolumn}
\usepackage{bm}

\usepackage{amsmath}
\usepackage{graphicx}
\usepackage{amsbsy,amsfonts}
\usepackage{amsthm}
\usepackage{natbib}
\usepackage{colonequals}
\usepackage[colorlinks]{hyperref}
\usepackage{hypcap}
\usepackage{enumitem}

\setlist[enumerate]{itemsep=0mm}

\newcommand{\ket}[1]{\left|#1\right\rangle}
\newcommand{\bra}[1]{\left\langle#1\right|}

\newcommand{\suppress}[1]{}

\def\squareforqed{\hbox{\rlap{$\sqcap$}$\sqcup$}}
\def\qed{\ifmmode\squareforqed\else{\unskip\nobreak\hfil
\penalty50\hskip1em\null\nobreak\hfil\squareforqed
\parfillskip=0pt\finalhyphendemerits=0\endgraf}\fi}

\newcommand{\eq}[1]{\hyperref[eq:#1]{(\ref*{eq:#1})}}
\renewcommand{\sec}[1]{\hyperref[sec:#1]{Section~\ref*{sec:#1}}}
\newcommand{\app}[1]{\hyperref[app:#1]{Appendix~\ref*{app:#1}}}
\newcommand{\fig}[1]{\hyperref[fig:#1]{Figure~\ref*{fig:#1}}}
\newcommand{\thm}[1]{\hyperref[thm:#1]{Theorem~\ref*{thm:#1}}}
\newcommand{\lem}[1]{\hyperref[lem:#1]{Lemma~\ref*{lem:#1}}}
\newcommand{\cor}[1]{\hyperref[cor:#1]{Corollary~\ref*{cor:#1}}}
\newcommand{\defn}[1]{\hyperref[def:#1]{Definition~\ref*{def:#1}}}

\DeclareMathOperator{\poly}{poly}
\DeclareMathOperator{\Cov}{Cov}

\newcommand{\T}{\mathrm{T}}

\newcommand{\expect}{\mathbb{E}}
\newcommand{\R}{\mathbb{R}}

\newcommand{\Hinv}{H_{-}}
\newcommand{\xinv}{\vec{x}_{-}}

\renewcommand{\eprint}[1]{\href{http://arxiv.org/abs/#1}{#1}}

\usepackage[bold]{hhtensor}
\usepackage{braket}
\usepackage{subcaption}
\input{Qcircuit}

\global\def \arxivmode {}

\ifx \arxivmode \undefined
\else
  
  \newcommand\arxivonly[1]{#1}
  \newcommand\prlonly[1]{}
\fi

\ifx \prlmode \undefined
\else
  
  \newcommand\arxivonly[1]{}
  \newcommand\prlonly[1]{#1}
  \renewcommand{\section}[1]{}
\fi

\begin{document}

\title{Hamiltonian Learning and Certification Using Quantum Resources}
\author{Nathan Wiebe}
\affiliation{Quantum Architectures and Computation Group, Microsoft Research, Redmond, WA 98052, USA}
\affiliation{Department of Combinatorics \& Optimization, University of Waterloo, Ontario N2L 3G1, Canada}
\affiliation{Institute for Quantum Computing, University of Waterloo, Ontario N2L 3G1, Canada}
\author{Christopher Granade}
\affiliation{Department of Physics, University of Waterloo, Ontario N2L 3G1, Canada}
\affiliation{Institute for Quantum Computing, University of Waterloo, Ontario N2L 3G1, Canada}
\author{Christopher Ferrie}
\affiliation{
Center for Quantum Information and Control,
University of New Mexico,
Albuquerque, New Mexico, 87131-0001}
\author{D. G. Cory}
\affiliation{Department of Chemistry, University of Waterloo, Ontario N2L 3G1, Canada}
\affiliation{Institute for Quantum Computing, University of Waterloo, Ontario N2L 3G1, Canada}
\affiliation{Perimeter Institute, University of Waterloo, Ontario N2L 2Y5, Canada}

\begin{abstract}
In recent years quantum simulation has made great strides culminating in experiments that operate in a regime that existing supercomputers cannot easily simulate.  Although this raises the possibility that special purpose analog quantum simulators may be able to perform computational tasks that existing computers cannot, it also introduces a major challenge: certifying that the quantum simulator is in fact simulating the correct quantum dynamics.  We provide an algorithm that, under relatively weak assumptions, can be used to efficiently infer the Hamiltonian of a large but untrusted quantum simulator using a trusted quantum simulator.  
We illustrate the power of this approach by showing numerically that it can inexpensively learn the Hamiltonians for large frustrated Ising models, demonstrating that quantum resources can make  certifying analog quantum simulators tractable.
\end{abstract}
\maketitle


Quantum information processing promises to dramatically advance physics and chemistry by providing efficient simulators for the Schr\"odinger or Dirac equations~\cite{lloyd_universal_1996,alan_qchem_2005,gerritsma_diracsim_2010}.
This is important because conventional methods are inefficient, scaling exponentially in the number of interacting subsystems.  Consequently, quantum simulations beyond a few tens of interacting particles are generally believed to be beyond the limitations of conventional supercomputers.   This inability to simulate large quantum systems means that important questions in condensed matter, such as the shape of the phase diagram for the Fermi--Hubbard model, remain open.
Analog quantum simulation raises the possibility that
special purpose \emph{analog devices} may be able to address such problems using current or near--future hardware~\cite{simon_simulation_2011,britton_simulation_2012,kim_simulation_2010}.  
A major objection to this avenue of inquiry is that analog simulators are not necessarily trustworthy~\cite{hauke_trust_2012,gogolin_boson_2013} and certification of them is not known to be efficient.
Without such certification, an analog simulator can at best only provide  hints about the answer to a given computational question.  A resolution to this problem is therefore essential if analog quantum simulators are to compete on an even footing with classical supercomputers.

An important first step towards a resolution is provided in~\cite{daSilva_practical_2011}, where it is shown that quantum systems with local time--independent Hamiltonians can be efficiently characterized given ensemble readout.  However, the method is not generally applicable, can be expensive and is not known to be either error robust or stable in cases where single shot measurements are used.
A number of machine learning and statistical inference methods  \cite{hentschel_machine_2010,hentschel_efficient_2011,sergeevich_characterization_2011,ferrie_how_2012,sergeevich_optimizing_2012,granade_robust_2012,lovett_differential_2013,svore_faster_2013} have been recently introduced to address similar problems in metrology or Hamiltonian learning.  In the context of Hamiltonian learning, such ideas have are known to be error--robust and lead to substantial reductions in the cost of high--precision Hamiltonian inference~\cite{granade_robust_2012}, albeit at the price of sacrificing the efficient scaling exhibited by~\cite{daSilva_practical_2011}.

\begin{figure}[t!]
\centering
\includegraphics[height=2.75in]{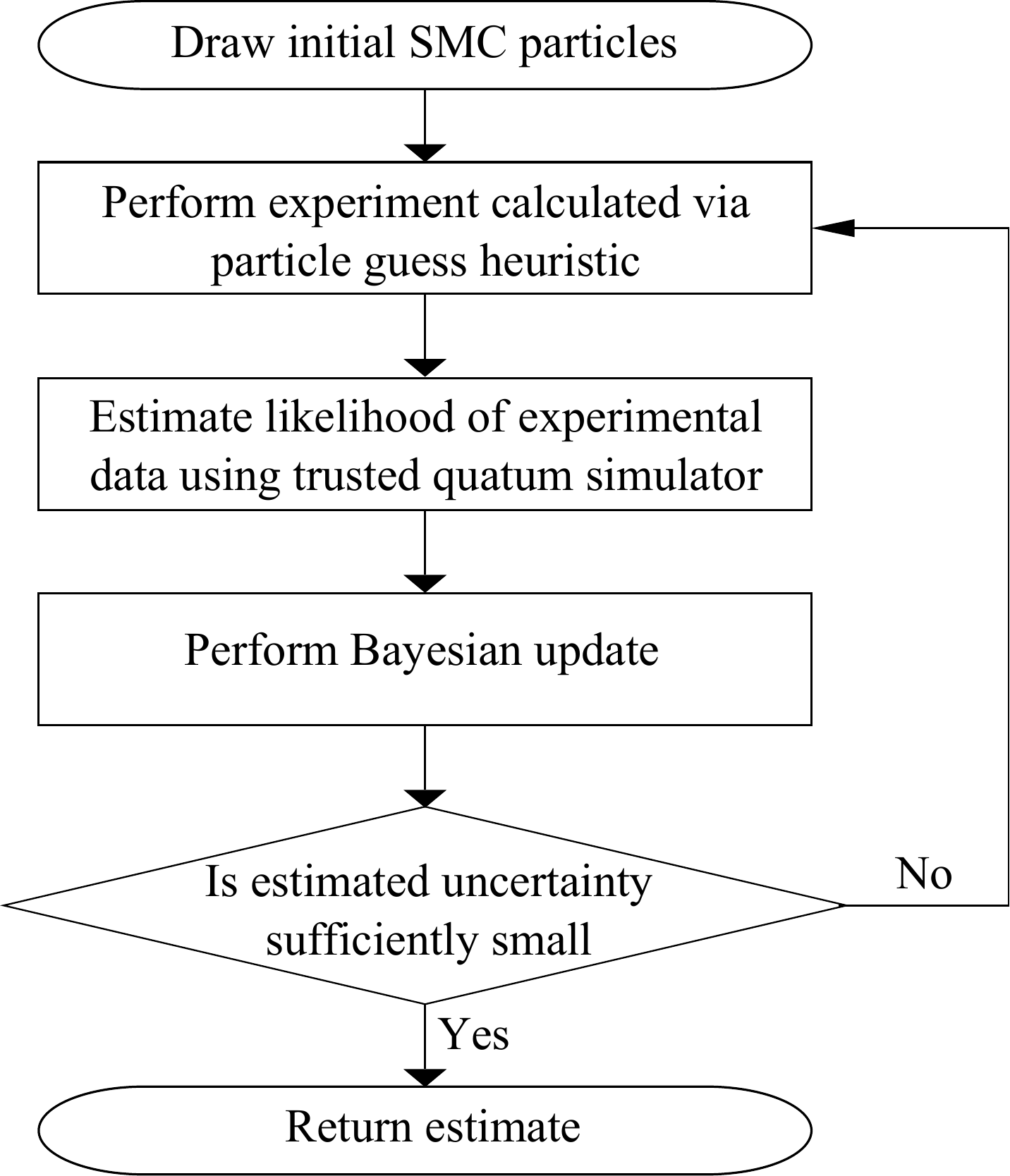}
\caption{\label{fig:flowchart} Flowchart for Hamiltonian learning algorithm.}
\end{figure}

We overcome these challenges by providing a robust method that can be used to characterize unknown Hamiltonians by unifying statistical inference with quantum simulation.  The key insight behind this is that Bayesian inference reduces the problem of Hamiltonian estimation to a problem in Hamiltonian simulation that can be efficiently solved using a trusted quantum simulator.  Our algorithm achieves this through the following steps. We begin by positing a Hamiltonian model for the system and a probability distribution over the parameters of the Hamiltonian model.  We then use a novel guess heuristic for the optimal experiment that adaptively chooses experiments based on the current uncertainty in the Hamiltonian.  The experiment is then performed and the trusted quantum simulator is used to efficiently compute the likelihood of the measurement outcome occurring if each hypothetical model were true.  These likelihoods are then used by the algorithm to update its knowledge of the Hamiltonian parameter via Bayes rule, resulting in an updated probability distribution, called the posterior distribution. This process is then repeated until the uncertainty in the unknown Hamiltonian parameters (as measured by the posterior variance) becomes sufficiently small.  This iterative process is depicted in \fig{flowchart}.




To make the problem concrete, we represent each hypothetical Hamiltonian $H_j$ by a vector of real numbers $\vec{x}_j \in \R^d$ such that $H_j = H(\vec{x}_j)$.  The Hamiltonian model is therefore specified by $H(\vec{x})$.  

We consider three classes of experiments that can be performed to infer the Hamiltonian, $H$, given an initial state $\ket{\psi}$ (typically a  pseudorandom state~\cite{emerson_pseudo-random_2003}): (a) Classical Likelihood Evaluation (CLE), (b) Quantum Likelihood Evaluation (QLE) and (c) Interactive Quantum Likelihood Evaluation (IQLE).  CLE is  the simplest of these experiments and is discussed in detail in~\cite{granade_robust_2012}.  It involves simply picking an experimental time $t$, and computing the likelihood $\Pr(D|\vec{x}_i)=|\bra{D}e^{-iH(\vec{x}_i)t}\ket{\psi}|^2$ using a classical computer, where $\vec{x}_i$ is a given set of Hamiltonian parameters and $D$ is the experimental outcome.  This function, known as the likelihood function, will not generally be efficiently computable on a classical computer because it involves quantum simulation.

In QLE experiments, a trusted quantum simulator is used to ameliorate these problems.  It does so by estimating $\Pr(D|\vec{x}_i)$ to be the fraction of times outcome $D$ occurs in a sufficiently large set of simulated experiments, which is efficient if $\Pr(D|\vec{x}_i)$ is only polynomially small.  
  This approach allows a complex quantum simulator, such as a fault tolerant quantum computer, to act as a certifier for an analog quantum simulator.  A trusted quantum simulator could also be constructed using a bootstrapping protocol wherein a smaller trusted analog simulator is the certifier.  This is possible if a compressed simulation scheme~\cite{kraus_compressed_2011} for the dynamics exists.

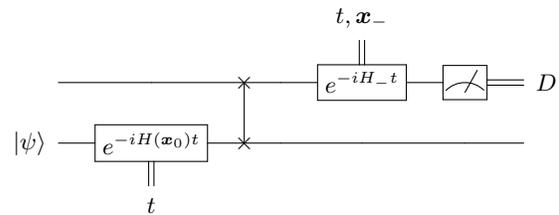
\begin{figure}
    \[
      \Qcircuit @R 1em @C 1.5em {
                        &                                      &                &                          & \ustick{t, \xinv}                 &        &                &  \\
                        & \qw                                  & \qswap \qwx[1] & \qw & \gate{e^{-i \Hinv t}} \cwx[-1] & \meter & \rstick{D} \cw &  \\
    \lstick{\ket{\psi}} & \gate{e^{-i H(\vec{x}_0) t}} \cwx[1] & \qswap         & \qw                      & \qw                                   & \qw    & \qw            &  \\
                        & \dstick{t}                           &                &                          &                                       &        &                &
      }
    \]
    \caption{\label{fig:models}IQLE. The upper register is the trusted simulator, while the lower register is the system under study.  QLE is conceptually similar but with $\Hinv=0$.}
\end{figure}

The Loschmidt echo famously shows that, for complex quantum systems, two nearly identical Hamiltonians will typically generate evolutions that diverge exponentially after a short time, before saturating at an exponentially small overlap~\cite{Haa06}.  This means that QLE will often be restricted to short evolution times  to guarantee efficiency (which is undesirable~\cite{granade_robust_2012}).  We resolve this by using IQLE experiments, which are described in~\fig{models}.  These experiments are reminiscent of the Hahn echo experiments commonly used in magnetic resonance and experimental quantum information processing~\cite{hahn_spin_1950}.
An IQLE experiment swaps the state of the unknown quantum system with that of a trusted quantum simulator then inverts the evolution based on a guessed Hamiltonian $\Hinv$.  The measurement in IQLE is always assumed to be in an orthonormal basis that has $\ket{\psi}$ as an element.  This produces $\Pr(D|\vec{x}_i)=|\bra{D}e^{i\Hinv t}e^{-iH(\vec{x}_i)t}\ket{\psi}|^2$.

Although the Loschmidt echo may also seem to be problematic for IQLE experiments, we exploit it in our guess heuristic for $\Hinv$.  We call this heuristic the ``particle guess heuristic'' (PGH), which chooses $\Hinv:= H(\xinv)$ by sampling $\xinv$ from the prior probability $\Pr({\vec{x}})$, which describes our current knowledge of the Hamiltonian parameters.  The set of parameters, $\xinv$, is called a \emph{particle} because it is described by a Dirac-delta distribution over parameter space.  The time $t$ is chosen by drawing a second particle $\xinv' \ne \xinv$ and setting $t=1/\|\xinv'-\xinv\|_2$.  As the uncertainty in the estimated parameter shrinks, the PGH adaptively picks longer times to ensure that informative experiments continue to be chosen as certainty about the unknown parameters increases.  The PGH also causes $e^{-iH(\vec{x}_i)t}$ to result in substantially different likelihoods for $\vec{x}_i$ that are within one standard deviation of the prior's mean, which we show in the appendix is optimal for certain learning problems.  

IQLE experiments with two outcomes also ensure that $\Pr(D|\vec{x})$ will not be exponentially small (with high probability) for $H$ an affine transformation acting on $\vec{x}$, since
\begin{align}
|\bra{\psi}e^{i\Hinv t}e^{-iHt}\ket{\psi}|&\ge 1-2\|H-\Hinv\|_2t \nonumber\\
&\ge 1- O(\|\vec{x}-\xinv\|_2t).
\end{align}
If the prior distribution has converged to a unimodal distribution centered near the correct Hamiltonian (this is typical for Bayesian inference of non-degenerate learning problems~\cite{granade_robust_2012}) then $\|\vec{x}-\xinv\|_2\in \Theta(1/t)$.  This means that if we use a POVM with two elements: $\ket{\psi}\!\!\bra{\psi}$ and its orthogonal compliment $\openone-\ket{\psi}\!\!\bra{\psi}$ then we expect (a) neither probability will be exponentially small if $\xinv$ and $\vec{x_j}$ are near the mean and (b) $\Pr(\psi|\vec{x_j})$ will typically be exponentially small for $H(\vec{x_j})$ that differ substantially from the correct Hamiltonian.  The PGH therefore leads to IQLE experiments that rapidly eliminate incorrect hypotheses about the correct Hamiltonian.


The measurement outcomes yielded by the experiments are \emph{immediately processed} using Bayesian inference, as described in~\fig{flowchart}.  This immediate processing allows our algorithm to adaptively choose experiments based on its current knowledge of the correct Hamiltonian.
The state of knowledge is represented by a distribution that is called, previous to the next update step, the \emph{prior}.  In the cases we consider, the initial prior distribution before any data is observed is taken to be uniform.  This encodes a state of maximum ignorance about the correct $\vec{x}$.
The prior distribution is updated as measurement outcomes are recorded using Bayes' rule, which gives the proper way of computing the probability of each $\vec{x}_j$ being correct given the observed data and the prior.  It states that if datum $D$ is recorded then
\begin{equation}
\Pr(\vec{x}_j|D) \propto \Pr(D|\vec{x}_j)\Pr(\vec{x}_j),\label{eq:bayes}
\end{equation}
up to a normalization factor and
 $\Pr(\vec{x}_j|D)$ is called the posterior distribution.  

Eq.~\eq{bayes} can be efficiently computed (for a polynomial number of $H_j$) only if the likelihood function $\Pr(D|\vec{x}_j)$
is tractable.  QLE and IQLE experiments allows $\Pr(D|\vec{x}_j)$ to be efficiently estimated, which removes the main obstacle to using Bayesian methods to learn the correct $\vec{x}$.


\begin{figure}[t!]
\includegraphics{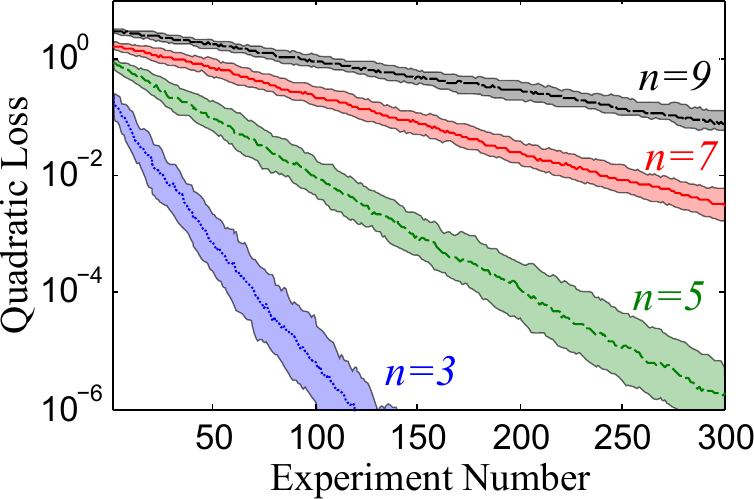}
\caption{The quadratic loss  plotted as a function of the number of inversion experiments for Ising models on the complete graph. The shaded areas show a $50\%$ confidence interval for the quadratic loss.}\label{fig:complete_scale}
\end{figure}

A secondary problem is that \emph{exact} computations of the update rule are intractable in practice because an infinite number of Hamiltonians  could potentially describe the system; hence, a probability
distribution over Hamiltonians cannot be exactly represented on either a classical or quantum computer.  This problem can be addressed by using the sequential Monte Carlo (SMC) approximation \cite{huszar_adaptive_2011,granade_robust_2012,doucet_tutorial_2011}, which approximates the probability distribution using a weighted sum of particles (Dirac delta functions). Each particle corresponds to a particlar $\vec{x}_j$ and is a hypothesis about the correct Hamiltonian parameters $\vec{x}$.  SMC assigns a weight $w_j$ to each particle that represents the probability of that hypothesis.  The weights are normalized such that $\sum_{j}w_j =1$.  The update rule for the probability distribution under the SMC approximation then becomes
$w_j \mapsto \Pr(D|\vec{x}_j)w_j$, followed by normalization.
If necessary, a resampling step is used after updating to ensure that the inference procedure remains stable, as discussed in~\cite{granade_robust_2012} and in \app{bayes}.

The algorithm then iteratively updates the weights $w_i$ and positions $\vec{x}_i$ of the sequential Monte Carlo particles representing the distribution over Hamiltonians $\Pr(H | D)$, conditioned on the data recorded at each step. In this way, the full state of knowledge at each step is iteratively carried forward, and is used to heuristically design future experiments according to the PGH.  Subsequent updates will then refine this estimate of the unknown Hamiltonian parameter until the uncertainty of the estimated Hamiltonian is sufficiently small, as measured by the trace of the posterior covariance matrix.

\begin{figure}[t!]
\centering
\includegraphics{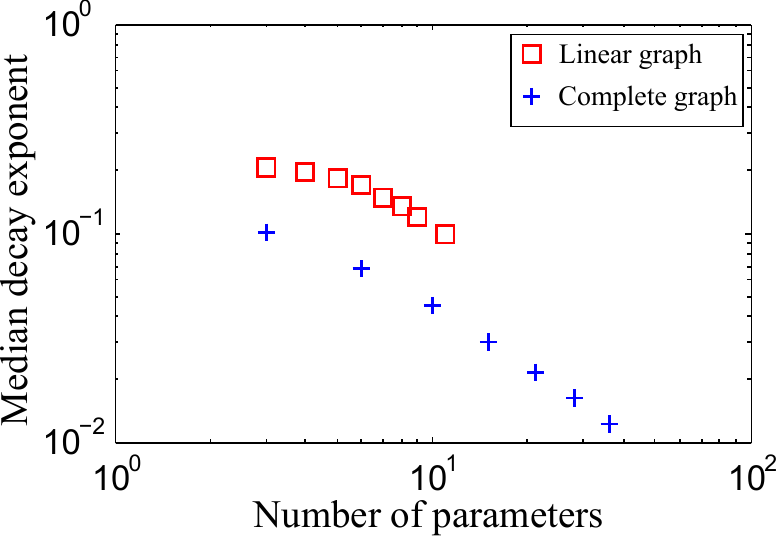}
\caption{The median decay exponent for the quadratic loss as a function of the number of parameters in the Ising model, $d$.}\label{fig:exp_scale}
\end{figure}

Now that we have discussed how our algorithm works, we will proceed to assess its cost.  We will show that the cost of Hamiltonian inference on a fixed number of IQLE experiments is exponentially smaller than the cost of using CLE.  This is significant because CLE  gives the best known methods for some problems~\cite{granade_robust_2012}.  

A natural measure of the cost is the number of quantum simulations needed to estimate the Hamiltonian parameters.  The total cost is therefore,
\begin{equation}
{\rm{Cost}} = N_{\rm steps}(\delta) \times {\rm{Cost}(\rm update;\epsilon)}.\label{eq:costbasic}
\end{equation}
Here $N_{\rm steps}$ is the number of updates needed to make the uncertainty less than $\delta$ and ${\rm{Cost}(\rm update;\epsilon)}$ is the number of samples from the trusted simulator that are needed to update the particle weights using Eq.~\eq{bayes} within error $\epsilon$ in the $1$--norm.
We show in \app{stability} that, with high probability,
${\rm{Cost}(\rm update;\epsilon)}$ scales as
\begin{equation*}
\frac{|\{\vec{x}_i\}|}{\epsilon^2}\left(\mathbb{E}_{D|H}\left[\frac{\max_k \Pr(D|\vec{x}_k)(1-\Pr(D|\vec{x}_k))}{\left( {\sum_k \Pr(D|\vec{x}_k)\Pr(\vec{x}_k)}\right)^2}\right]\right).
\end{equation*}
This implies that the update process will be efficient if the number of particles required is small and the resultant probability distribution is not too flat.  That is, $|\{\vec{x}_i\}|\in O({\rm poly}(n))$ and  $\sum_k \Pr(D|H(\vec{x}_k))\Pr(H(\vec{x}_k))\in O(1/{\rm poly}(n))$, where $n$ is the number of interacting systems.  It has been shown that SMC algorithms require a number of particles that scales sub--exponentially in $d$~\cite{beskos_stability_2011}, which itself may not be a function of $n$.  This means that in practice, a small number of particles will typically be required.  The robustness of the algorithm to sampling errors is discussed in~\cite{ferrie_likelihood-free_2013} as well as in \app{stability}, so relatively large $\epsilon$ can be tolerated.




If the posterior distribution has converged to a unimodal distribution such that $\vec{x}$ is within a fixed distance from the mean, then the PGH and~\eq{bayes} ensure that $\mathbb{E}_{\Hinv} [|\!\bra{\psi}e^{i\Hinv t}e^{-iHt}\ket{\psi}|^2] \in \Theta(1)$ since $t\in \Theta(|\vec{x}-\xinv|^{-1})$.  If a two outcome measurement is used then  Markov's inequality implies that  $\sum_k \Pr(D|H(\vec{x}_k))\Pr(H(\vec{x}_k))\in \Theta(1)$ with high probability.  
By a trivial generalization of this argument, it is clear that a super--polynomial reduction in the cost of performing~\eq{bayes} relative to CLE is obtained with high probability for IQLE experiments if $d\in O({\rm poly}(n))$ and the  \emph{effective} number of outcomes, $\sum_j \Pr(j|\vec{x}_k)^{-2}$, is at most $O({\rm poly}(n))$ for each $\vec{x}_k$.

In contrast, QLE experiments may not lead to a super--polynomial separation in the cost estimates for generic Hamiltonians and large $t$ because $\sum_k \Pr(D|H(\vec{x}_k))\Pr(H(\vec{x}_k))\in 2^{-\Theta(n)}$ with high probability for complex quantum systems~\cite{ududec_equilibration_2013,Haa06}.  This  can be rectified by choosing small $t$ as per~\cite{daSilva_practical_2011}, but such QLE experiments will be much less informative~\cite{granade_robust_2012}.



If a fixed number of updates are required, then the previous discussion and~\eq{costbasic} suggest that IQLE will provide an exponential advantage over CLE.  If inference within a fixed error tolerance, $\delta$, is required then the cost estimate is much more challenging.
 Each two-outcome measurement yields at most one bit of information about $H$ per measurement hence $N_{\rm steps}(\delta) \in \Omega(d\log_2(1/\delta))$.  For most models of interest, $d$ is polynomial (or even constant) in $n$ and hence a small number of updates should typically suffice.  It is, however, unclear whether this lower bound is tight; hence, we turn to numerical evidence to show that our algorithm can efficiently learn Hamiltonians in certain cases.



Consider the problem of learning $H({\vec{x}})$ using IQLE experiments for an Ising model with no transverse field:
\begin{equation}\label{eq:Ising}
 H(\vec{x}) = \sum_{(i,j)\in G} x_{i,j}\ \sigma_z^{(i)}\sigma_z^{(j)},
\end{equation}
where $G$ is the edge set of an interaction graph on $n$ qubits. Unless otherwise specified, we take $x_{i,j}\in [-1/2,1/2]$ uniformly at random.  We take the initial state for the evolution to be $\ket{\psi}=\ket{+}^{\otimes n}$.  We choose this Hamiltonian not only because it is physically relevant~\cite{richerme_trapped-ion_2013}, but also for numerical expediency, since the learning process require the algorithm to perform thousands of simulated evolutions of the initial state.  All measurements are performed in the eigenbasis of $X^{\otimes n}$.  Restricting the measurements to two outcomes is unnecessary for these experiments because IQLE and the PGH concentrates $\Pr(D|\vec{x}_i)$ over a small number of outcomes for this Hamiltonian.

\fig{complete_scale} shows that the quadratic loss (a generalization of the mean--squared error for multiple parameters) shrinks exponentially with the number of experiments performed; however, the rate at which the error decreases slows as the number of qubits $n$ increases.  This is expected because $d=n(n-1)/2$ for the case of a complete interaction graph, which implies that the learning problem becomes more difficult as $n$ increases.  The data for interactions on the line is similar and is presented in \app{line} and QLE data is given in \app{QLE}.

\begin{figure}[t!]
\centering
\includegraphics[width=0.95\linewidth]{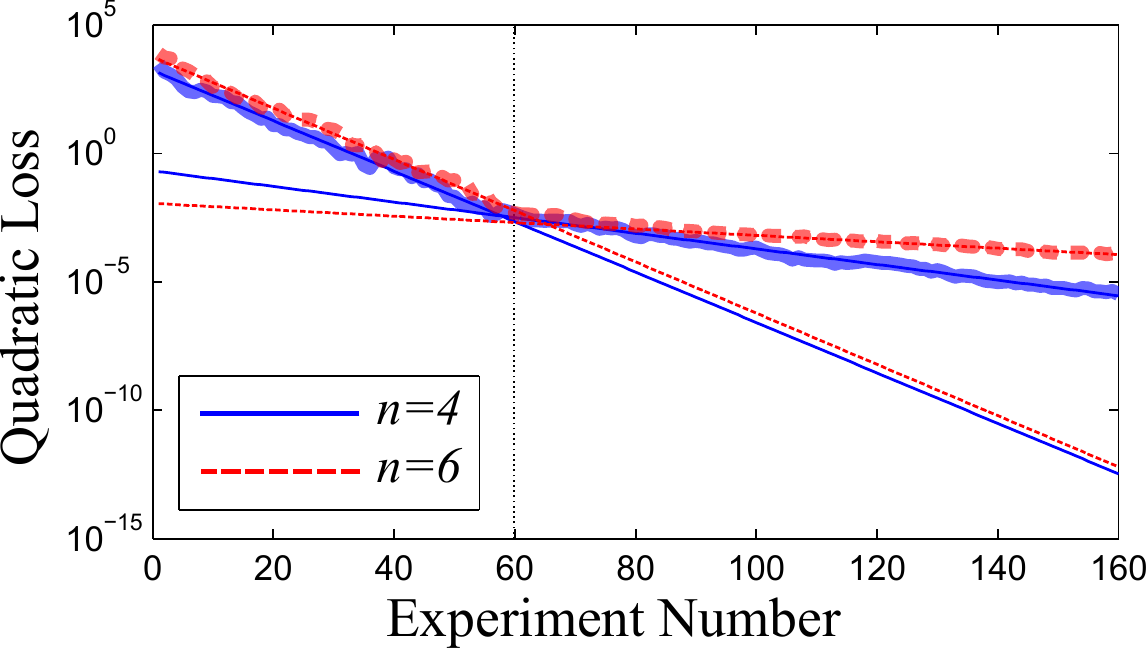}
\caption{An approximate $1$ parameter Ising model on the complete graph.  The thin lines give the best fits to the exponential decays, which scale as $e^{-0.07N }$ and $e^{-0.23N}$ for $n=4$ and as $e^{-0.029N}$ and $e^{-0.23N}$ for $n=6$ qubits.\label{fig:corner}}
\end{figure}

The rate at which the learning process slows as $n$ increases is investigated in
\fig{exp_scale}.  We examine the slowing of the learning problem by fitting the quadratic loss, $\delta$, in each experiment to $Ae^{-\gamma N_{\rm steps}}$.  The median decay exponent, which is the median of the values of $\gamma$ attained for a set of experiments with constant $n$, measures how rapidly the algorithm learns the unknown parameters. \fig{exp_scale}  shows that these decay constants scale as $O(1/d)$ for the complete graph, and provides weaker evidence for the line.  This implies that $N_{\rm steps}(\delta)=O(d\log(1/\delta))$ for this Hamiltonian, which implies that the inference is efficient.  Similarly, the PGH implies that the total simulation time needed (for fixed $|\{\vec{x}_i\}|$) scales as $N_{\rm steps}\delta^{-1}\approx \delta^{-3}$, which is relevant in cases where the cost of a simulation is dominated by the evolution time.

Although $d=n(n-1)/2$ or $d=n-1$ in the examples considered above, $d$ can be approximately independent of $n$ in some cases.
An example of this behavior is given in \fig{corner}, where we consider the case where each of the $x_{i,j}$ is approximately the same value chosen uniformly on $[0,100]$, but with small normally distributed fluctuations with mean $0$ and variance $10^{-4}$.  This causes the learning problem to be effectively one--dimensional initially, and then transition to $d=n$ when the small fluctuations  need to be identified to learn the Hamiltonian parameters within a fixed accuracy.  The transition from a single-- to a multi--parameter learning problem happens at $\delta \approx d\times 10^{-4} \approx 10^{-3}$, which coincides with the point when the slope in \fig{corner} changes.  This emphasizes that the cost of Hamiltonian estimation using our method only implicitly depends on $n$ through $d$.  In fact, the difference in the observed scaling of $\gamma$ is approximately a factor of $2.5$, which is what would be expected if  $\gamma\propto 1/d$.


In conclusion, we have shown that Bayesian inference combined with the SMC approximation provides an ideal way to leverage a (potentially non--universal) quantum
simulator to characterize an unknown or unreliable quantum system.  We provide theoretical evidence that shows that the update rule, which is at the heart of the learning algorithm, can be performed efficiently using quantum resources.  We then illustrate the practicality of the algorithm and show that it is capable of learning unknown Ising couplings with surprisingly few experiments even in the presence of sampling errors.  We will show elsewhere that the algorithm is highly resilient to depolarizing noise and other forms of noise that can be introduced via a noisy swap gate.

\begin{acknowledgements}
  We acknowledge Troy Borneman for suggesting bootstrapping and also Krysta Svore and Allan Geller for useful feedback and discussion.  The numerical experiments performed here used SciPy, F2Py and QInfer~\cite{SciPy2001,peterson_f2py:_2009,ferrie_qinfer_2012}.  This work was supported by funding from USARO-DTO, NSERC, CERC and CIFAR.  CF was supported in part by NSF Grant Nos.~PHY-1212445 and PHY-1005540.
\end{acknowledgements}




\onecolumngrid
    \appendix
\section{Supplemental Data}\label{app:supp}
\subsection{Error Scaling for Linear Interaction Graph}\label{app:line}
In the main body of the text, we showed that our algorithm learns information about the Hamiltonian at a rate that scales exponentially with the number of experiments taken for both the complete graph and the line, but only presented an example of the raw data for the case where the interaction graph is complete.  For completeness, we provide here analogous data for the case where the interaction graph is a line.

\begin{figure}[t!]
\includegraphics[width=0.5\columnwidth]{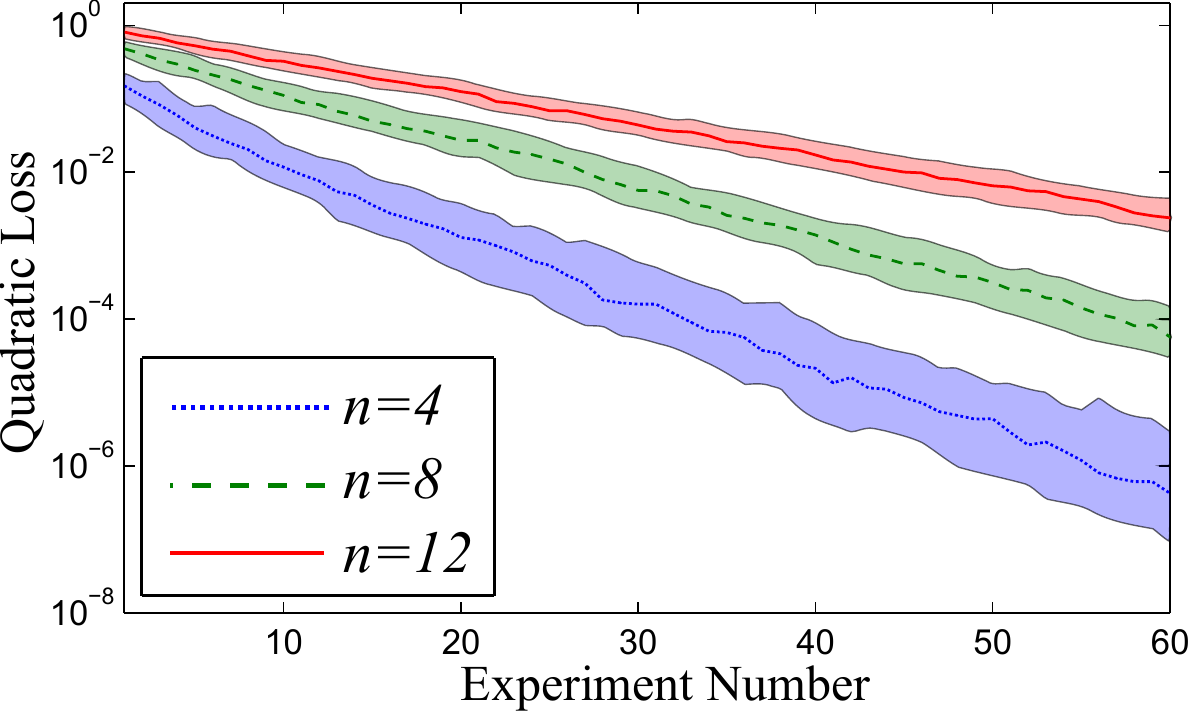}
\caption{This figure shows the quadratic loss plotted as a function of the number of IQLE experiments for $4,8,12$ qubits (from bottom to top) interacting on the line. The dashed lines show a $50\%$ confidence interval for the quadratic loss. $10~000$, $10~000$ and $20~000$ particles were used in the $n=4$, $n=8$ and $n=12$ cases respectively.}\label{fig:line_scale}
\end{figure}

This data clearly shows that IQLE experiments are similarly effective in the case of linear interaction graphs in that the data follows an exponential scaling.  The fits of the median quadratic loss to an exponential of the form $Ae^{-\gamma N}$, where $N$ is the experiment number, is  given in \fig{exp_scale}.

\subsection{Error Scaling for QLE}\label{app:QLE}
A major problem facing the use of QLE experiments is efficiently estimating the likelihood function using quantum simulation.  Despite this  problem, if we grant the algorithm the ability to do perfect likelihood evaluations at unit cost then QLE experiments can be highly informative.  For example, the typical variation of the likelihood function for  QLE experiments with large $t$ on random Hamiltonians acting on $n$ qubits drawn from the Gaussian Unitary Ensemble (such random Hamiltonians model complex quantum systems with time-reversal symmetry~\cite{Haa06}) is on the order of $2^{-n}$~\cite{ududec_equilibration_2013} which is on the same order as the typical values of the likelihood.  This means that, if \emph{we do not consider} sampling errors, then late time QLE experiments will allow learning to occur even for complex Hamiltonians.

In spite of this, Hamiltonian learning using QLE experiments is expected to be much less stable in this regime.  This is because $\Pr(D|\vec{x}_j)$ can be approximately the same as (or larger than) $\Pr(D|\vec{x})$ even if $\|\vec{x}_k -\vec{x}\|_2$ is large (here $\vec{x}$ is the correct Hamiltonian parameter).  This can cause the learning algorithm to get confused and move particles to near $\vec{x}_k$ during the resampling step.  This makes it harder for the algorithm to recover from the bad inference and continue to learn.  Thus even if we grant QLE experiments the ability to perform \emph{exact} likelihood evaluations at unit cost, then we still do not expect such experiments to be as robust to bad inferences as IQLE.

\begin{figure}[h!]
\includegraphics[width=0.5\columnwidth]{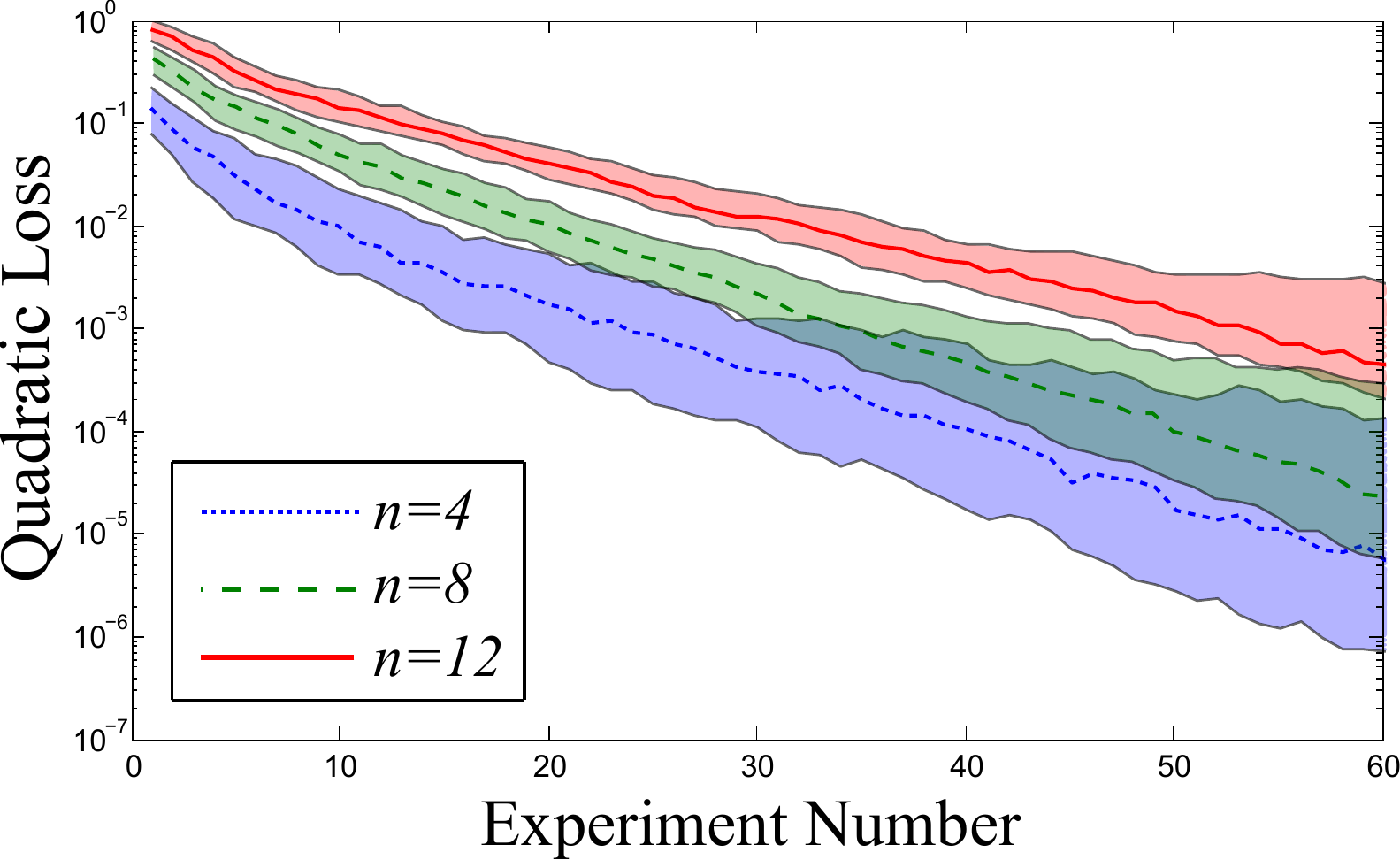}
\caption{The quadratic loss plotted as a function of the number of QLE experiments for $4,8,12$ qubits (from bottom to top) interacting on the line. The dashed lines show a $50\%$ confidence interval for the quadratic loss. $10~000$, $10~000$ and $20~000$ particles were used in the $n=4$, $n=8$ and $n=12$ cases respectively.}\label{fig:QLEline_scale}
\end{figure}

\fig{QLEline_scale} confirms these expectations.  It shows that the $25^{\rm th}$ percentile of the quadratic loss for QLE experiments for the Ising model on the line is similar to that of IQLE experiments.  The most notable difference between the data sets is that the $50\%$ confidence intervals overlap.  This is because, in each case, the learning algorithm is more likely to get confused in IQLE experiments versus QLE experiments.  Additionally, the $75^{\rm th}$ percentile of the quadratic loss is \emph{much worse} for $n=12$ and suggests that the learning algorithm eventually fails in such cases.  Similar problems are observed in the median and $75^{\rm th}$ percentile of the $n=4$ data.  These problems are not fatal: they can be addressed by repeating the learning algorithm several times and using a majority vote scheme to reduce the impact of instances where the learning algorithm becomes confused.  As mentioned previously, we expect real problems to emerge for QLE experiments when inexact likelihood calls are considered.

\subsection{Errors in Likelihood Evaluations}
The numerical examples provided in the main body of the text assumed that the error in the inference process due to using a finite number of samples to compute the likelihoods $\Pr(D|\vec{x}_i)$ is negligible.  Here we provide evidence showing that the learning process is robust to such errors for IQLE experiments on the Ising models considered previously.


In particular, let us define $\mathcal{P}$ to be the uncertainty in the estimated probability that results from estimating the likelihood in using the trusted quantum simulator.  Here we simulate the use of MLE (Maximum Likelihood Estimation) or ALE (Adaptive Likelihood Estimation \cite{ferrie_likelihood-free_2013}) methods by adding normally distributed noise with zero mean and standard deviation $\mathcal{P}$, and then clip the likelihood to the interval $[0,1]$.  This is chosen in preference to MLE or ALE because it is expedient to compute and it models the results of either method closely.  We find that even if $\mathcal{P}$ is a large constant then our algorithm continues to reduce the quadratic loss at a rate that scales as $e^{-\gamma N}$; albeit at a reduced value of $\gamma$.
This clearly indicates that we do not need to take $\epsilon$ to small if we need small error.

\begin{figure}[!]
\centering
\includegraphics[width=0.5\columnwidth]{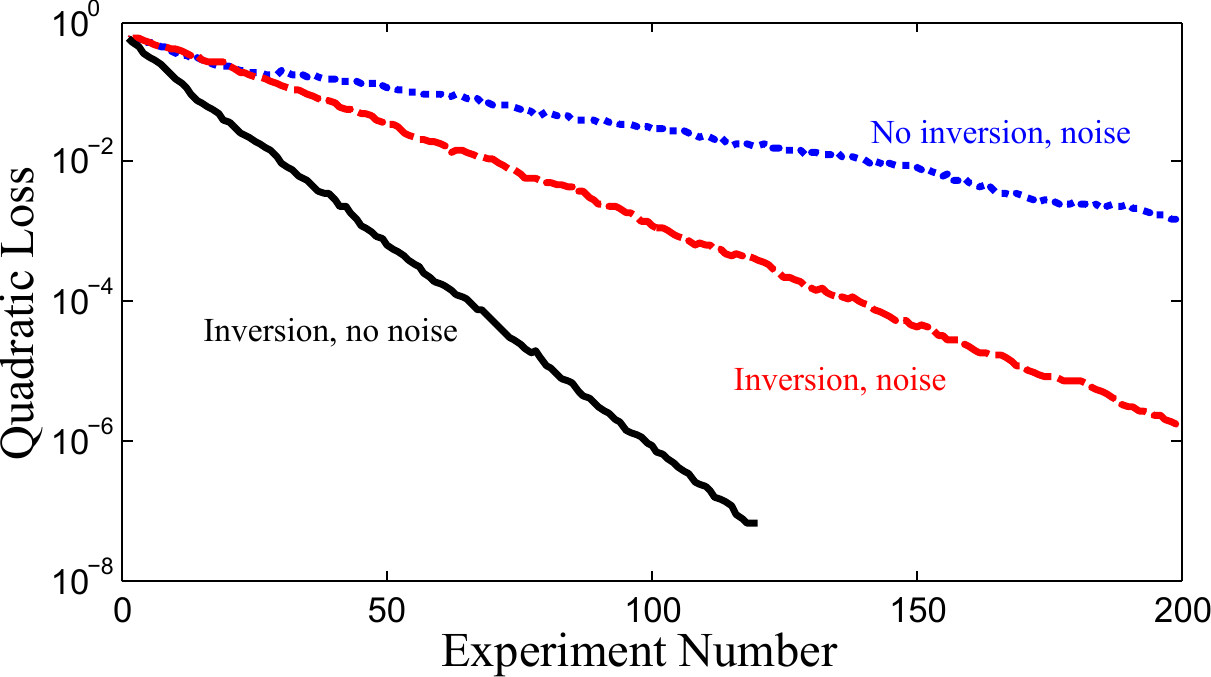}
\caption{This plot shows the median quadratic loss for a $9$ qubit Ising model on the line for the cases where inversion is used and when inversion is not used for cases where the sample standard deviation is $\mathcal{P}$ is the uncertainty in the estimated likelihood. 10~000 particles were used for the calculation.}\label{fig:poison}
\end{figure}

The robustness of Bayesian inference using the SMC approximation is illustrated in \fig{poison}, where we show that our algorithm is robust to sampling errors for $9$ qubits interacting on the line for $\mathcal{P}=0.1$ and $\mathcal{P}=0$.  We see that the data for QLE experiments agrees with that of IQLE experiments for short times, which is expected because the probability distribution has not had time to reach its maximum support.  At later times, QLE exeperiments with $\mathcal{P}=0.1$ fare much worse than IQLE experiments.  Nonetheless, IQLE experiments (and QLE for this value of $\mathcal{P}$) still exhibit exponential scaling of the error with the number of experiments.  This may be surprising because the errors in $\Pr(D|\vec{x}_i)$ can be as large as $0.1$, which one may assume would be catastrophic given that many of the outcomes will have probability less than $0.1$ in such models.  We note that in particular, IQLE experiments are more robust to such noise than QLE experiments.  This is because the inversion employed by IQLE concentrates the probability over a smaller number of outcomes; leading to smaller relative errors in the likelihood evaluations in such cases.

\begin{figure}[t!]
\centering
\includegraphics[width=0.5\columnwidth]{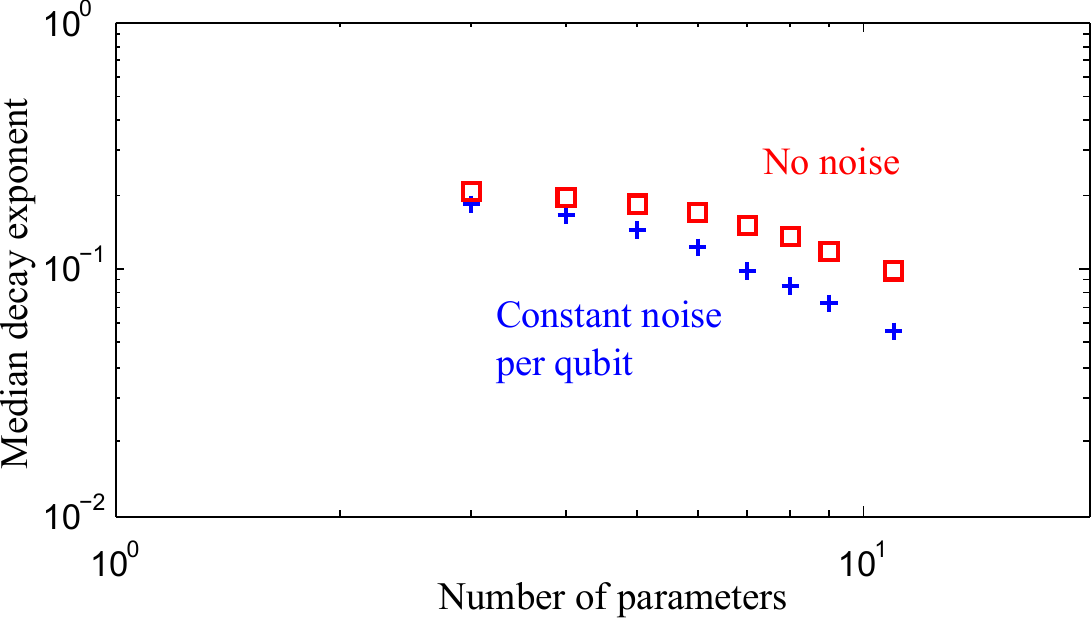}
\caption{This plot shows the median value of $\gamma$ computed for the case of IQLE experiments where the interaction graph is a line and $n$ ranges from $1$ to $12$ with different levels of noise.  20~000 particles were used for these numerical experiments.}\label{fig:poisonscale}
\end{figure}

\fig{poisonscale} gives a more clear picture of the effects of sampling error on the resultant distribution for IQLE experiments.  We observe that the presence of such noise does not qualitatively change the scaling of $\gamma$, where $\gamma$ is the decay exponent that describes how the quadratic loss shrinks as more experimental data is provided.  Specifically, we find that for $\mathcal{P}=0$, $\gamma\propto d^{-1}$ whereas for $\mathcal{P}=0.4/n$ (which corresponds to $\epsilon \approx \sum_{i=1}^d 0.4/n \approx 0.4$) we find that $\gamma\propto d^{-3/2}$.  This shows that large sampling errors do not necessarily prevent our algorithm from learning the Hamiltonian parameters at a rate that scales as $Ae^{-\gamma N}$ and further suggests that this learning process is efficient for the problem of learning unknown Ising couplings.  It also suggests that a constant value of $\epsilon$ may suffice for certain experiments.

The surprising robustness of our method comes in part from the fact that the likelihood function must be approximated for each particle.  This means that if the algorithm errs in the update of a particular particle due to inexact evaluation of the likelihood function then it may not err substantially in evolving the total probability in the region that many such particles are in.  For example, consider a region $R$ that has 10~000 particles in it.  The probability density in that region is then $\sum_{\vec{x}_i \in R} w_i/V(R)$, where $V(R)$ is the volume of the region.  We then see that if errors are independent and identically distributed over each particle, then the total error in the update of the probability density will be roughly $1/100$ the error that would be expected if all the errors were in fact correlated.  Thus the  robustness of the algorithm may be understood in part as a consequence of the fact that the errors are (approximately) unbiased about the true likelihood and that the particle number will typically be large for high precision inferences.

\section{Bayesian Inference of Hamiltonians}\label{app:bayes}

Sequential Monte Carlo (SMC) has before been considered in the context of quantum information \cite{huszar_adaptive_2011}, and in particular for its utility in estimating Hamiltonian dynamics \cite{granade_robust_2012}.
Here, we summarize and review the sequential Monte Carlo algorithm and approximation, as SMC is an important tool for the practical implementation of statistical inference according to Bayes' rule, and in particular for our proposed methods.   The SMC approximation is of central importance here because Hamiltonian models are typically parameterized by a vector of real numbers rather than discrete numbers.  This means that there are an infinite number of hypothetical Hamiltonians that could represent the system, which makes the update of the prior distribution intractable.  The SMC approximation is used to model the continuous distribution over model parameters (which is computationally difficult to sample from) by a discrete distribution
 that preserves the low--order moments of the distribution and thereby making estimation of the unknown Hamiltonian parameters tractable.

To clarify, suppose we have fixed an input state $\ket\psi$ and measurement basis $\{\ket{D}\}$, but that the Hamiltonian under which the state evolves is unknown.
Had we known that the Hamiltonian was $H$, we apply Born's rule to obtain the probability distribution for the outcomes of the experiment:
\begin{equation}
\Pr(D|H)=|\bra{D}e^{-iHt}\ket{\psi}|^2.
\end{equation}
This is called the \emph{likelihood function}.  When we write a probability distribution $\Pr(\vec{x}|y)$, we are stating how likely the proposition $\vec{x}$ is true \emph{given} $y$ is known to be true.  In a Hamiltonian learning problem, $H$ is unknown and the measurement result is given.
Bayes' rule provides a way to invert the conditioning to provide the probability that $H$ is the true Hamiltonian given that datum $D$ is recorded:
\begin{equation}
\Pr(H|D) =\frac{\Pr(D|H)\Pr(H)}{\Pr(D)}=\frac{\Pr(D|H)\Pr(H)}{\int \Pr(D|H)\Pr(H)\,\mathrm{d} H}.
\end{equation}
Here, $\Pr(H)$ is called the \emph{prior} and formally encodes any \emph{a priori} knowledge of the unknown Hamiltonian.  The probability of interest, $\Pr(H|D)$ is called the \emph{posterior} since it encodes our \emph{a posteriori} knowledge.  The final term $\Pr(D)$ can simply be thought as a normalization factor that can be found implicitly by integrating over the unnormalized distribution.
Since each measurement is statistically independent given $H$, the processing of the data can be done on- or off-line; Bayesian updating (or Bayesian learning or Bayesian inference) allows us to sequentially update our knowledge of the Hamiltonian through a sequence of probability distributions $\Pr(H|\{D_1, D_2,\ldots\})$.

In practice, the Bayesian update rule and the expectations listed above are analytically and computationally intractable since they involve integrals over multidimensional parameter spaces.  However, if we drop the requirement of a deterministic algorithm, we can efficient compute them using Monte Carlo techniques.  Our numerical algorithm fits within the subclass of Monte Carlo methods called \emph{sequential Monte Carlo} or SMC \cite{doucet_tutorial_2011}.  

The first step in the approximation method is to think of $H$ as a function that maps a parameterization $\vec{x}$ of a Hamiltonian to a Hermitian operator $H(\vec{x})$.
Doing so allows us to reduce the dimension of the random variable that we are reasoning about, called the model dimension, by using knowledge about the class
of Hamiltonians which are plausible given the physics of the system.
We then approximate the probability distribution by a weighted sum of Dirac delta-functions,
\begin{equation}
\Pr(H(\vec{x})) \approx \sum_{j=1}^{|\{\vec{x}_i\}|} w_j  \delta(\vec{x} - \vec{x}_j),
\end{equation}
where the weights at each step are iteratively calculated from the previous step via
\begin{equation}
w_j \mapsto \Pr(D|\vec{x}_j) w_j,
\end{equation}
followed by a normalization step.  The elements of the set $\{\vec{x}_j\}_{j=0}^{|\{\vec{x}_i\}|}$ are called \emph{particles}.  Here, ${|\{\vec{x}_i\}|}$ is the number of particles and controls the accuracy of the approximation.  Like all Monte Carlo algorithms, the SMC algorithm approximates expectation values, such that
\begin{equation}
\mathbb{E}_{\vec{x}}[f(H(\vec{x}))] \approx \sum_{j=1}^{|\{\vec{x}_i\}|} w_j f(H(\vec{x}_j)).
\end{equation}
In other words, sequential Monte Carlo allows us to efficiently compute multidimensional integrals with respect to the measure defined by the probability distribution.

An iterative numerical algorithm such as SMC requires care to ensure stability.  In the next section, we derive the conditions for stability of the algorithm.  But first we describe one additional, and important, step in the iteration.  The step is called \emph{resampling} and is required to ensure that the SMC particles explore the space of Hamiltonians rather than staying fixed at the ${|\{\vec{x}_i\}|}$ initially chosen hypotheses.  This is necessary both intuitively and, as we will see next, computationally.

The idea is simple: if the weight associated to a particle is too is small, move the particles to a region where the weight is large.  We follow the methodology of Liu and West \cite{Liu2000Combined}.  First, to determine when to resample, we compare the effective sample size $N_{\text{ess}} = 1/\sum_j w_j^2$ to a threshold (typically $|\{\vec{x}_i\}|/2$). If the threshold is not met, we randomly select ${|\{\vec{x}_i\}|}$ new particles according to the distribution of the current weights. Additionally, we incorporate randomness to search larger volumes of the parameter space.  This randomness is inserted by
applying a random perturbation to the location of each new particle.  Thus, the new particles are randomly spread around the previous locations of the old.
After drawing ${|\{\vec{x}_i\}|}$ new particles, we set the weight of each new particle to $1/{|\{\vec{x}_i\}|}$ so that $N_{\text{ess}} = |\{\vec{x}_i\}|$.  To clarify, the Liu and West resampler algorithm updates the position of a particle $\vec{x}_i$, which is sampled from the posterior distribution, by drawing a particle from a Gaussian distribution with mean
\begin{equation}
\vec{\mu}_i = a\vec{x}_i + (1-a)\vec{\mu},
\end{equation}
where $\vec{\mu}=\mathbb{E}_{\vec{x}} [\vec{x}]$ is the posterior mean of the particle location and $a\in[0,1]$ is a constant.  The covariance matrix for the Gaussian distribution is given by
\begin{equation}
\vec{\Sigma}=(1-a^2)\Cov(\vec{x}),
\end{equation}
where $\Cov(\vec{x})$ is the covariance matrix for the particle positions.  The resampler therefore introduces randomness into the problem that depends on the current level of uncertainty in the unknown Hamiltonian parameters.  We find for the learning problems that we consider $a=0.9$ performs well, whereas for simpler learning problems $a=0.98$ was found to be superior~\cite{granade_robust_2012,Liu2000Combined}.
Full algorithmic details of SMC, including resampling, are given in \cite{granade_robust_2012}.

The resultant posterior probability provides a full specification of our knowledge.  However, in most applications, it is sufficient---and certainly more efficient---to summarize this distribution.   In our context, the optimal single Hamiltonian to report is the mean of the posterior distribution (here, we have omitted for brevity the fact that the posterior is conditioned on the data)
\begin{equation}
\mu_H := \mathbb{E}_{\vec{x}}[H(\vec{x})] = \sum_{j=1}^{|\{\vec{x}_i\}|}w_j H(\vec{x}_j).
\end{equation}
However, a single point is the space of unknown Hamiltonians does not provide information of the uncertainties in this estimate.  For that we turn to \emph{regions}.  In particular, the set of Hamiltonians $X$ is an $\alpha$-credible region if
\begin{equation}
\Pr(H\in X) \geq 1-\alpha.
\end{equation}
That is, a set is an $\alpha$-credible region if no more than $\alpha$ probability mass exists outside the region or, equivalently, at least $1-\alpha$ probability mass is contained within the region.

After a sufficient number of experiments, we assume the posterior distribution will be approximately Gaussian in terms
of our chosen parameterization, so that $\vec{x} \sim \mathcal N(\hat{\vec{x}},\Cov[\vec{x}])$, where $\hat{\vec{x}} = \expect[\vec{x}]$.  Then, an $\alpha$-credible region estimate is given by the covariance ellipse
\begin{equation}
X = \{H(\vec{x}): (\vec{x}-\mathbb{E}_{\vec{x}}[\vec{x}])^{\T}\Cov[\vec{x}]^{-1}(\vec{x}-\mathbb{E}_{\vec{x}}[\vec{x}]) \leq Z_\alpha^2\},
\end{equation}
where $Z_\alpha^2$ is the $\alpha$-quantile of the $\chi_d^2$ distribution.  Such estimates are important because they allow SMC methods to characterize the uncertainty in an estimate of the unknown Hamiltonian~\cite{granade_robust_2012}.  We do not emphasize the ability of our learning algorithm to perform region estimation in the main body of the text, but the algorithm's capability of specifying the uncertainty in the unknown Hamiltonian through the form of a region estimation provides a powerful advantage over tomographic methods wherein such a characterization of the uncertainty is much less natural.

\section{Solution in tractable cases}
\label{app:derivations}

The mathematical tools we use to solve the Hamiltonian identification problem are those of decision theory and statistical learning.  We have used a combination of methods from computation statistics to approximate the optimal solution to the problem of learning a Hamiltonian.  However, for the case of a single unknown parameter, the equations can be solved analytically.  These solutions provide much of the insight into designing the numerical algorithm to solve the general problem as well as serve to explain, in a broader context, why our method succeeds.

\subsection{Statistical decision theory of learning}
To evaluate the performance of any algorithm, we compare the estimated Hamiltonian parameters $\hat{\vec{x}}$ to the true parameters $\vec{x}_0$ by using the
quadratic loss $L(\hat{\vec{x}}, \vec{x}) = \|\hat{\vec{x}} - \vec{x}\|^2 $.  This loss function generalizes the mean squared error to multiple parameters, and quantifies the error we incur due to the estimation procedure.

Our task is to choose an \emph{estimator} $\hat{\vec{x}}(D)$, a function from the possible data sets to valid parameters.  This problem is most naturally cast in the language of decision theory.  There is ostensibly one general approach: minimize---in some sense---the expected loss, or \emph{risk}.  That is we choose the estimator which satisfies
\begin{equation}\label{eq:opt Bayes mean loss}
\hat{\vec{x}}_{\rm opt} := \operatorname{argmin}_{\hat{\vec{x}}}\mathbb E_{\vec{x},D}[\|\vec{x}-\hat{\vec{x}}(D)\|^2],
\end{equation}
where the expectation is with respect to the distribution of $\vec{x}$ and $D$ for the given experiment.  This objective function is denoted $r(\vec{e})$ for the experiment $\vec{e}$, and called the \emph{Bayes risk}.  Under some regularity conditions, the unique best strategy is the Bayesian one, selecting as the estimator the mean of the posterior distribution
\begin{equation}
\hat{\vec{x}}_{\rm opt} = \mathbb E_{\vec{x}|D}[x].
\end{equation}
An important and useful consequence of using the quadratic loss is that the Bayes risk is equal to the expected trace of the covariance matrix of the posterior distribution (in the case of a single parameter it is simply the variance).

\subsection{Single parameter problem}

For the single parameter problem, the Hamiltonian reads
\begin{equation}
H(x) = x \sigma_z^{(1)}\sigma_z^{(2)},
\end{equation}
and the initial state is $\ket +$ and final measurement is in the basis $\{\ket +,\ket -\}$  (labeling the outcomes $\{0,1\}$.  If we evolve for a time $t$ and allow an IQLE experiment with inversion Hamiltonian $\Hinv:=H(\xinv)$,  the output probability distribution (the likelihood function) is
\begin{equation}
\Pr(d|x;\xinv,t) = \frac12(1 + (1-2d)\cos[2(x-\xinv)t])
\end{equation}
This model, for CLE experiments, was studied in detail in \cite{ferrie_how_2012}.  To obtain asymptotic expressions, we assume that probability distribution of $x$ is approximately Gaussian and remains so after a subsequent measurement.  The risk incurred between these two measurements then provides an asymptotic approximation to Bayes risk of the algorithm.

Formally, we assume
\begin{equation}
\Pr(x|\mu,\sigma) = \frac{1}{\sqrt{2\pi\sigma^2}}\exp\left(-\frac{(x-\mu)^2}{\sigma^2}\right),
\end{equation}
with mean $\mu$ and variance $\sigma^2$.  The posterior distribution
\begin{equation}
\Pr(x|d, \mu, \sigma; \xinv,t) = \frac{\Pr(d|x,\mu,\sigma;\xinv,t)\Pr(x|\mu,\sigma)}{\Pr(d|\mu,\sigma;\xinv,t)}
\end{equation}
gives a mean of
\begin{equation}
\hat{x}_{\rm opt} = \mu + \frac{2i(1-2d)\sigma^2 t(e ^{4i\mu t}-e^{4i\xinv t})}{2 e^{2t(i(\mu+\xinv)+\sigma^2 t)} + (1-2d)(e ^{4i\mu t}-e^{4i\xinv t})},
\end{equation}
which is the final estimator.  The risk incurred by this estimator (which, recall, is the optimal one) is explored next.

\subsection{Asymptotic risk and the particle guess heuristic}
\begin{figure}\centering
   \includegraphics[width=1\columnwidth]{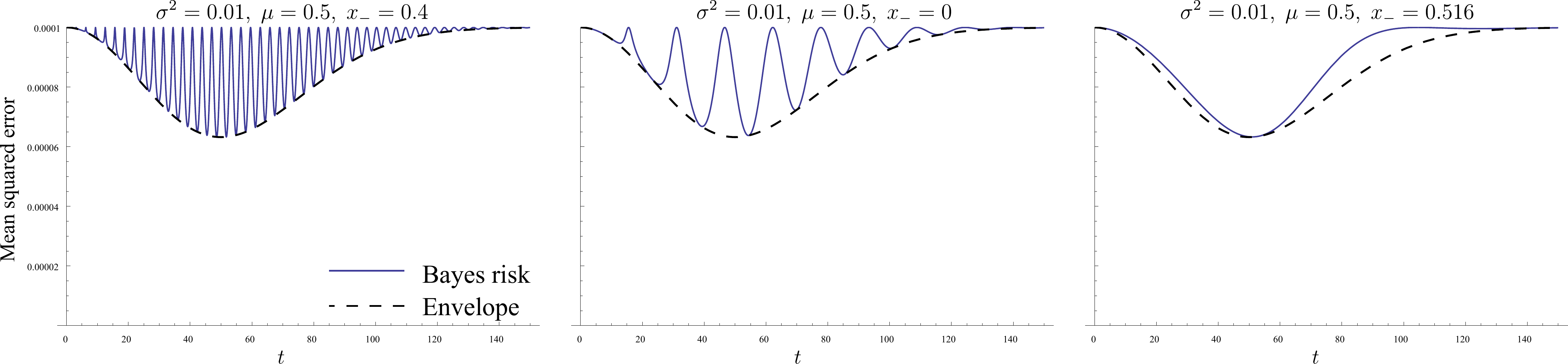}
  \caption{\label{fig:risk_final} The mean squared error of the optimal estimator as a function of evolution time.  The initial distribution of $x$ is normal $\mathcal N(0.5,0.01)$.  The left figure shows the mean squared error with no inversion.  The remaining figures show the mean squared error with an inversion Hamiltonian.  Note that the right figure uses $\xinv \approx \mu + \sigma$.  The black dotted line lower bounds the mean squared error and its minimum value is the ultimate limit on the performance of any strategy.}
\end{figure}

These calculations rapidly become too cumbersome to display.  The behavior of these more complex systems, however, can be described concisely with a few graphs, as shown in \fig{risk_final}.

\fig{risk_final} shows the mean squared error for different choices of $(t,\xinv)$.  Notice the envelope
\begin{equation}
1-4\sigma^2 t^2 e^{-4\sigma^2t^2}\leq \frac{r(\xinv,t)}{\sigma^2} \leq 1.
\end{equation}
This tells us that the posterior variance cannot increase on average.  In other words, there is no such thing as a strictly bad experiment.  It also gives us a theoretical lower bound; the ``risk envelope'' has a minimum at
\begin{equation}
t_{\rm opt} = \frac{1}{2\sigma}.
\end{equation}
This leads to
\begin{equation}
\min_{\xinv,t} r(\xinv,t) = (1-e^{-1})\sigma^2.
\end{equation}
That is, per measurement, the risk is reduced by a factor of about $0.63$, which leads to the exponential scaling observed in \cite{ferrie_how_2012}.

Notice, however, in \fig{risk_final} that the risk rapidly oscillates within the risk envelope.  This shows, in particular, that the minimum corresponds to the solution of challenging global optimization problem.  This is where we see the advantage of inversion;  the effect of inversion is to ``wash out'' these oscillations.  Thus, errors in approximations misplacing the optimal evolution time are less severe.  Based on numerical testing, the optimal experimental inversion parameter is
\begin{equation}\label{eq:exp design 1D}
\xinv \approx \mu \pm \sigma.
\end{equation}

These results and conclusions are only valid for the 1-dimensional parameter estimation problem.   However, from these results we can gain intuition for what ought to happen in the multi-dimensional case.  First, since the role of time is identical, we would expect that the optimal algorithm achieves exponential scaling, as we indeed observe.  Second, we should expect that the optimal time for each experiment be proportional to some function of the inverse covariance matrix of the current distribution.  Computing and inverting the covariance matrix is computationally inconvenient and so we use the heuristic
\begin{equation}
t = \frac{1}{\|\vec{x}' - \vec{x}\|},
\end{equation}
where $\vec{x}\neq \vec{x}'$ are two particles drawn at random from the distribution of particles weights.  This is a proxy for the inverse of the standard deviation.  Finally, a computationally efficient analog of the experiment design in equation \eq{exp design 1D} is to simply select $\xinv$ at random from the distribution of particle weights.  As we will see next, this added randomization has a positive effect when additional errors are present.

One final note before we move on is that the above analysis assumes the distribution is approximately Gaussian.  This will eventually be true but in practice we require a ``warm-up'' phase of experiment designs before we employ the the heuristics motivated by the asymptotic analysis.  Fortunately, the randomization included in the particle guess heuristic provides a way to adaptively warm-up the learning algorithm without including an ad--hoc warmup heuristic, as was done in previous studies~\cite{granade_robust_2012}.
\subsection{Robustness of inversion to sampling error}

\begin{figure}\centering
   \includegraphics[width=1\columnwidth]{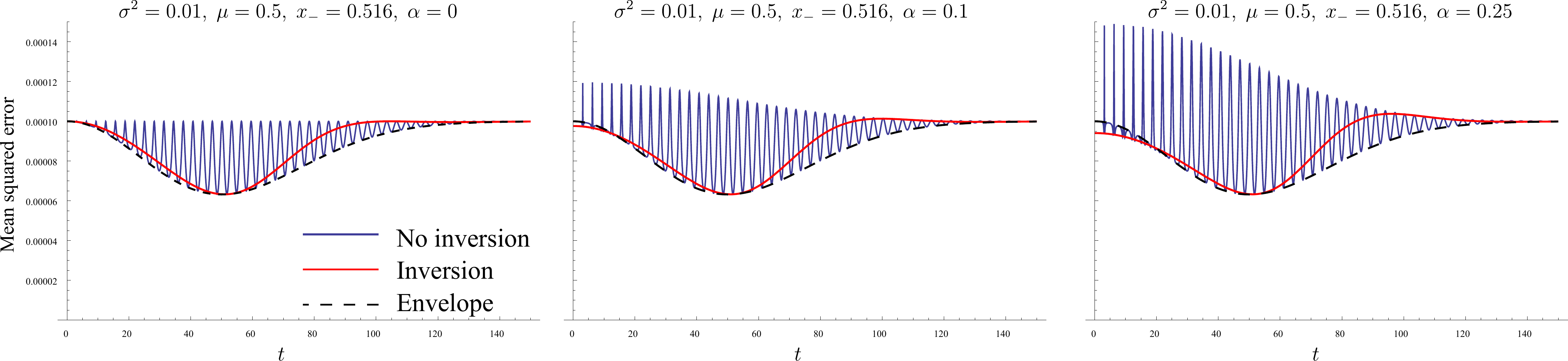}
  \caption{\label{fig:risk_noise_final} The mean squared error of the optimal estimator as a function of evolution time.  This is identical to the far right plot in \fig{risk_final} except that we had added an additional bit-flip noise source of varying strength.}
\end{figure}

For a two-outcome model, the only possible errors (regardless of origin---physical, modeling, sampling, etc.) manifest as bit-flips.  If we assume the process is symmetric, we have a noisy version of the likelihood function,
\begin{equation}
\Pr(d|x;\xinv,t,\alpha) = \alpha + (1-2\alpha)\Pr(d|x;\xinv,t),
\end{equation}
where $\alpha$ is the probability of a bit-flip.  Now, since we assume the the algorithm is blind to this added noise, the posterior does not change.  Thus, the estimator (the posterior mean) and the variance do not change either.  If this seems odd, one must think of the posterior as a logical construct which is updated with assumed model---not the true model.  To evaluate the Bayes risk however, we must take the average with respect to data of true model:
\begin{equation}
r(\xinv,t,\alpha) = \mathbb E_{d|\xinv,t,\alpha}[{\rm Var}_{x|d;\xinv,t}(x)].
\end{equation}
This quantity is shown in \fig{risk_noise_final} for various values of $\alpha$.  The important thing to note is that the strategy with no inversion possesses a risk which can now \emph{increase}.  Now ``bad'' experiments are just as likely as good ones near the optimal evolution time.  Remarkably, the inversion model is complete insensitive to any strength of noise near the optimal evolution time.  Moreover, the ``particle guess heuristic'' achieves the same performance independent of noise, which implies that the experimenter need not change their strategy depending on whether noise is present or not.

\subsection{Consistency in multiple dimensions}

The above analysis considered the case of a single unknown parameter.  While this makes the statistical lessons learned equally valid when moving to more unknown parameters plausible, it would be more comforting to have similar results for more than a single parameter.  Unfortunately, the integrals required appear to be analytically intractable.  We can, however, perform simulations to obtain an approximate function form for the Bayes risk.  To this end, we consider the 3-qubit problem:
\begin{equation}
H(x_{1},x_2) = x_1 \;\sigma_z^{(1)}\sigma_z^{(2)}+x_2\;\sigma_z^{(2)}\sigma_z^{(3)}.
\end{equation}

\begin{figure}\centering
   \includegraphics[width=.95\columnwidth]{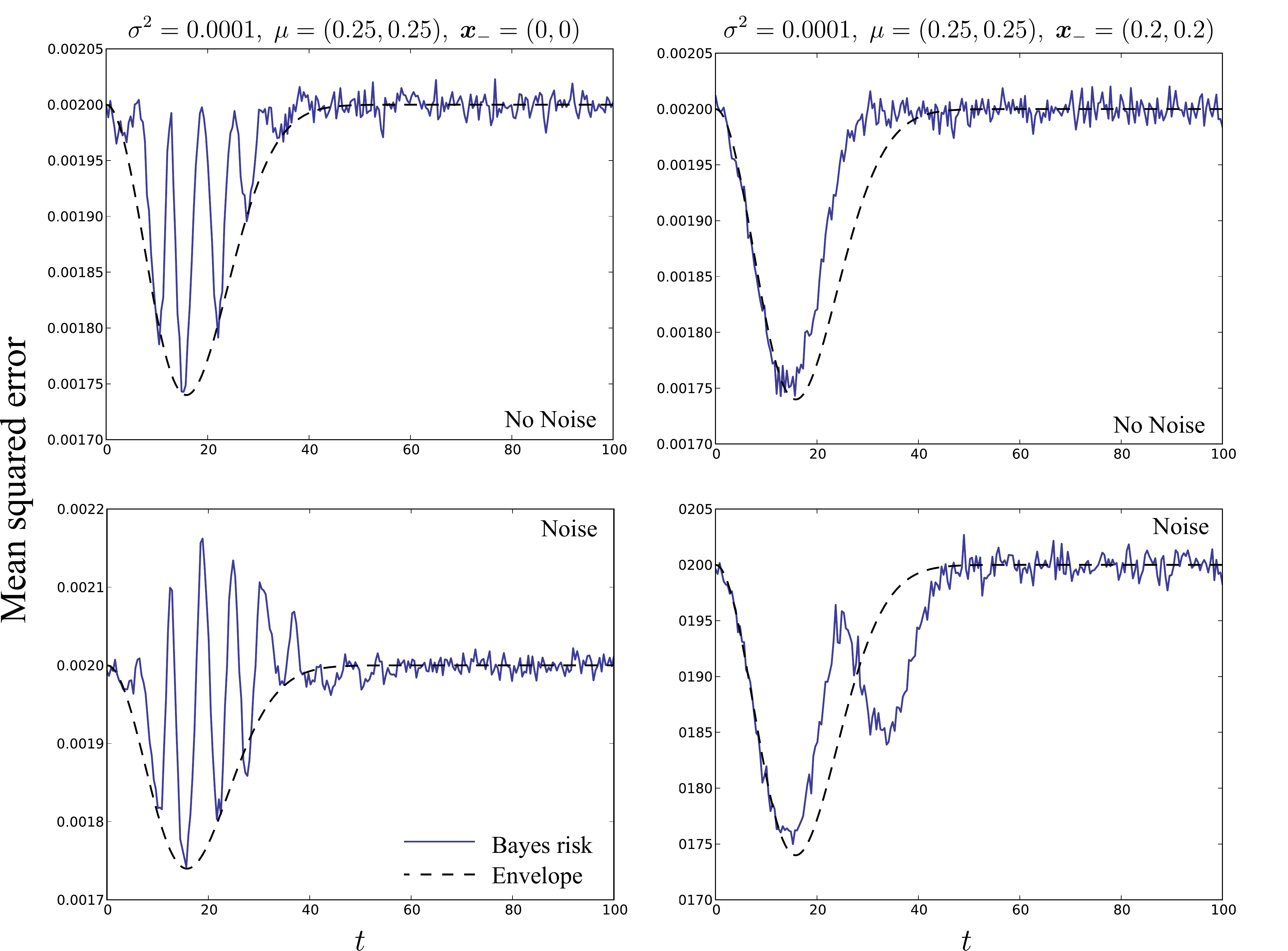}
  \caption{\label{fig:risk_2d_final} The mean squared error of the optimal estimator as a function of evolution time.  The initial distribution of $(x_1,x_2)$ is chosen to be normally distributed with mean $\mu = (0.25,0.25)$ and diagonal covariance matrix with equal variances for both coordinates: $\sigma^2 =0.001$. The top row displays the mean squared error when there is no noise with the left plot showing the case of no inversion and right showing the choice of an inversion Hamiltonian with parameters roughly a distance from the mean that is given by the square root of the trace of the covariance matrix ($2\sigma$ in this case).
The lower two plots show the mean squared error for the same two inversion strategies when there is 10\% noise present.}
\end{figure}

The results of the simulations, analogous to those presented in Figures \ref{fig:risk_final} and \ref{fig:risk_noise_final}, are shown in \fig{risk_2d_final}.  The conclusions drawn from the 2-qubit case remain; inversion enhances the performance of the estimation algorithm by smoothing out the Bayes  risk and leaving the improvement unchanged near the optimal evolution time.

\section{Conditions for Asymptotic Stability of Bayesian Inference}\label{app:stability}

We have already discussed the need for resampling as a means of maintaining the stability of performing Bayesian inference using the SMC approximation.  Here we discuss why these instabilities arise, and whether there are other sources of instabilities that can arise in quantum Hamiltonian estimation.  We show that the errors in the updating procedure will, on average, be small given that experiments are chosen that do not yield small likelihoods for probable events and given that the particle weights used in the SMC approximation do not become too small.

We consider, for now, only one step in the updating procedure.  There are two sources of errors that can arise in the update procedure: (1) errors in the prior that have arisen due to previous approximate updates or numerical errors in the initial prior (2) errors in the likelihood evaluation.  Let us assume that datum $D$ is obtained and let the error--free prior probability of Hamiltonian $H(\vec{x}_j)$, for any $j$, be denoted $\Pr(\vec{x}_j)$ and similarly the actual likelihood is $\Pr(D|\vec{x}_j)$.  We then denote the approximate analogs of these distributions ${\Pr}(\vec{x}_j)$ and $\tilde{\Pr}(D|\vec{x}_j)$.  The error in the posterior probability of $\vec{x}_j$ is

\begin{equation}
\epsilon_j =\left|\frac{\Pr(D|\vec{x}_j)\Pr(\vec{x}_j)}{\sum_j \Pr(D|\vec{x}_j)\Pr(\vec{x}_j)}-\frac{\tilde \Pr(D|\vec{x}_j)\tilde \Pr(\vec{x}_j)}{\sum_j \tilde \Pr(D|\vec{x}_j)\tilde \Pr(\vec{x}_j)} \right|
\end{equation}
For simplicity, we will now introduce variables that describe the variation of the approximate probabilities from the precise probabilities.  These deviations can, in many circumstances, be thought of as random variables since the majority of the error
in this protocol will arise from using sampling to estimate the likelihood function.
\begin{align}
\tilde \Pr(\vec{x}_j)&:=\eta'_j + \Pr(\vec{x}_j)\nonumber\\
\tilde \Pr(D|\vec{x}_j)&:=\eta_j + \Pr(D|\vec{x}_j)
\end{align}
We make use of the fact that
\begin{equation}
\eta  := \sum_j \Pr(D|\vec{x}_j)|\eta'_j| +\Pr(\vec{x}_j)|\eta_j| +|\eta_j\eta'_j| \le \frac 1 2 \sum_j \Pr(D|\vec{x}_j)\Pr(\vec{x}_j).
\end{equation}

Then assuming that $\max\{|\eta'_j|,|\eta_j|\}\in O(\eta)$ we find by using Taylor's theorem and the triangle inequality that
\begin{align}
\epsilon_j &= \left|\frac{ [\Pr(D|\vec{x}_j)+\eta_j] [\Pr(\vec{x}_j)+\eta'_j]}{\sum_k [\Pr(D|H(\vec{x}_k))+\eta_k][\tilde \Pr(H(\vec{x}_k))+\eta'_k]}-\frac{\Pr(D|\vec{x}_j)\Pr(\vec{x}_j)}{\sum_k \Pr(D|H(\vec{x}_k))\Pr(H(\vec{x}_k))} \right|\nonumber\\
&\le\Biggr|(\Pr(\vec{x}_j)+\eta'_j)(\Pr(D|\vec{x}_j)+\eta_j)\left(\frac{1}{\sum_k \Pr(D|H(\vec{x}_k))\Pr(H(\vec{x}_k))}+\frac{2\eta}{(\sum_k \Pr(D|H(\vec{x}_k))\Pr(H(\vec{x}_k)))^2} \right)\nonumber\\
&\qquad\qquad-\frac{\Pr(D|\vec{x}_j)\Pr(\vec{x}_j)}{\sum_k \Pr(D|H(\vec{x}_k))\Pr(H(\vec{x}_k))} \Biggr|\nonumber\\
&\le\left(\frac{\Pr(D|\vec{x}_j)|\eta'_j|+\Pr(\vec{x}_j)|\eta_j|}{\sum_k \Pr(D|H(\vec{x}_k))\Pr(H(\vec{x}_k))}+\frac{2\eta \Pr(D|\vec{x}_j)\Pr(\vec{x}_j)}{(\sum_k \Pr(D|H(\vec{x}_k))\Pr(H(\vec{x}_k)))^2} \right)+O(\eta^2)\label{eq:epsilonj}
\end{align}
The overall error as measured by the $1$--norm is $\epsilon = \sum_j \epsilon_j$ and hence~\eq{epsilonj} gives
\begin{align}
\epsilon&\le\frac{3\sum_k \Pr(D|H(\vec{x}_k))|\eta'_k|+\Pr(H(\vec{x}_k))|\eta_k| }{\sum_k \Pr(D|H(\vec{x}_k))\Pr(H(\vec{x}_k))}+O(\eta^2)\nonumber\\
&\le\frac{3\left(\sqrt{\sum_k \Pr^2(D|H(\vec{x}_k))}\sqrt{\sum_k |\eta'_k|^2}+ \sqrt{\sum_k \Pr^2(H(\vec{x}_k))}\sqrt{\sum_k |\eta_k|^2} \right)}{\sum_k \Pr(D|H(\vec{x}_k))\Pr(H(\vec{x}_k))}+O(\eta^2).\label{eq:error}
\end{align}

Equation \eq{error} provides an upper bound for the error in the Bayesian update for a fixed measured datum $D$.  In practice, surprising outcomes can destabilize the update according to~\eq{error}.  The contribution of such surprising results to the overall error is small if
\begin{align}
\sqrt{\sum_j \eta_j'^2}&\ll \frac{\sum_k \Pr(D|H(\vec{x}_k))\Pr(H(\vec{x}_k))}{\sqrt{\sum_k \Pr^2(D|H(\vec{x}_k))}},\label{eq:delta'}\\
\sqrt{\sum_j \eta_j^2}&\ll \frac{\sum_k \Pr(D|H(\vec{x}_k))\Pr(H(\vec{x}_k))}{\sqrt{\sum_k \Pr^2(H(\vec{x}_k))}}=\sqrt{N_{\rm ess}}\left({\sum_k \Pr(D|H(\vec{x}_k))\Pr(H(\vec{x}_k))}\right),\label{eq:delta}
\end{align}
where $N_{\rm ess}$ is the effective sample size.

These equations give two different criteria for the stability of the Bayesian update.  Equation~\eq{delta'} states that if the weights of the particles are too small then an unreasonably small value of $\eta'_j$ may be required to ensure that the error in the update is small.  This justifies the need for using resampling in SMC methods, and further justifies the criteria used for resampling in our algorithm: $N_{\rm ess}= (\sum_k \Pr^2(H(\vec{x}_k)))^{-1}\le |\{\vec{x}_i\}|/2$.  Equation~\eq{delta} makes a more interesting claim.   It states that the update rule can become unstable if the expectation value of $\Pr(D|H(\vec{x}_k))$ over the prior $\Pr(H(\vec{x}_k))$ is small for typical values of $D$.

Eqns.~\eq{error} and~\eq{delta} imply that the error due to estimating the likelihood via sampling is asymptotically bounded above by $\epsilon$ if
\begin{equation}
\sum_j \eta_j^2\in O\left( \epsilon^2N_{\rm ess}\left({\sum_k \Pr(D|H(\vec{x}_k))\Pr(H(\vec{x}_k))}\right)^2\right).\label{eq:deltaAssBd1}
\end{equation}
Since ${\sum_j \eta_j^2}\le |\{\vec{x}_i\}| \eta_{j\max}^2$, where $\eta_{j\max}=\max_j \eta_j$, we have that~\eq{deltaAssBd1} is satisfied if
\begin{equation}
\eta_{j\max}\in O\left( \epsilon\sqrt{\frac{N_{\rm ess}}{|\{\vec{x_i}\}|}}\left({\sum_k \Pr(D|H(\vec{x}_k))\Pr(H(\vec{x}_k))}\right)\right).\label{eq:deltaAssBd2}
\end{equation}
Our criteria for resampling is that $N_{\rm ess} \le |\{\vec{x}_i\}|/2$.  So, we have that $N_{\rm ess}\in \Theta(|\{\vec{x}_i\}|)$ and hence~\eq{deltaAssBd1} is implied by
\begin{equation}
\eta_{j\max}\in O\left( \epsilon\left({\sum_k \Pr(D|H(\vec{x}_k))\Pr(H(\vec{x}_k))}\right)\right).\label{eq:deltaAssBd3}
\end{equation}

We use as our estimate of $\Pr(D|H(\vec{x}_k))$ the fraction of samples drawn from the simulator that yield outcome $D$.
The resultant distribution for the number of samples that yield $D$ is a binomial distribution with mean $N_{\rm samp} \Pr(D|H(\vec{x}_k))$ and variance $N_{\rm samp}\Pr(D|H(\vec{x}_k))(1-\Pr(D|H(\vec{x}_k)))$.
Hence, if $N_{\rm samp}$ samples are drawn from the simulator then the uncertainty in our estimate of $\Pr(D|H(\vec{x}_k))$ obeys
\begin{equation}
\eta_{j\max}\in O\left(\sqrt{ \frac{\max_k \Pr(D|H(\vec{x}_k))(1-\Pr(D|H(\vec{x}_k)))}{{N_{\rm samp}}}}\right).
\end{equation}
Therefore we have from~\eq{deltaAssBd3} that~\eq{deltaAssBd1} is satisfied if
\begin{equation}
\sqrt{ \frac{\max_k \Pr(D|H(\vec{x}_k))(1-\Pr(D|H(\vec{x}_k)))}{{N_{\rm samp}}}}\in O\left( \epsilon{\sum_k \Pr(D|H(\vec{x}_k))\Pr(H(\vec{x}_k))}\right),
\end{equation}
which is equivalent to saying that
\begin{equation}
N_{\rm samp} \in \Omega\left(\frac{\max_k \Pr(D|H(\vec{x}_k))(1-\Pr(D|H(\vec{x}_k)))}{\left( \epsilon{\sum_k \Pr(D|H(\vec{x}_k))\Pr(H(\vec{x}_k))}\right)^2} \right),
\end{equation}
We require that $N_{\rm samp}$ samples are drawn for each particle in $\{\vec{x}_i\}$ and hence it is sufficient to take a number of simulations that scales as
\begin{equation}
N_{\rm sim} \in \Theta\left(\frac{|\{\vec{x}_i\}|\max_k \Pr(D|H(\vec{x}_k))(1-\Pr(D|H(\vec{x}_k)))}{\left( \epsilon{\sum_k \Pr(D|H(\vec{x}_k))\Pr(H(\vec{x}_k))}\right)^2} \right).
\end{equation}

Our method uses the mean of the posterior distribution as an estimator for the true Hamiltonian, $H(\vec{x})$, which means that more work is needed to determine how an error of $\epsilon$ in the update procedure propagates to errors in the mean and the variance of the posterior distribution.  Let $\mu_H:=\sum_i \Pr(H(\vec{x}_i)|D)H(\vec{x}_i)$ be the posterior mean and $\tilde\mu_H:=\sum_i \tilde\Pr(H(\vec{x}_i)|D)H(\vec{x}_i)$ be the posterior mean calculated by approximate likelihood evaluation.
The error in the estimated Hamiltonian, as measured by the $2$--norm is then
\begin{equation}
\|\mu_H -\tilde\mu_H\|\le \max_i \|H(\vec{x}_i)\| \sum_i |\Pr(H(\vec{x}_i)|D)-\tilde{\Pr}(H(\vec{x}_i)|D)|= \max_i \|H(\vec{x}_i)\| \epsilon.
\end{equation}
Similarly, it is straight forward to see that
\begin{equation}
\left|\sum_{i}\Pr(H(\vec{x}_i)|D) \|H(\vec{x}_i)-\mu_H\|^2-\tilde\Pr(H(\vec{x}_i)|D) \|H(\vec{x}_i)-\tilde\mu_H\|^2\right|\in O( \max_i \|H(\vec{x}_i)\|^2\epsilon),
\end{equation}
where $\sum_{i} \Pr(H(\vec{x}_i)|D) \|H(\vec{x}_i)-\mu_H\|^2$ is the posterior variance.

It may be tempting to conclude that after $N$ steps, the error in the estimate is $N\max_i \|H(\vec{x}_i)\| \epsilon$, but because the Bayesian update rule is non-linear it is difficult to prove such a bound.  Instead, note that Bayesian inference is robust to the choice of prior~\cite{granade_robust_2012} and thus the inference process will remain stable under such errors.  We therefore can consider beginning the inference process using the erroneous posterior as the prior and expect convergence if the relative errors in the variance are small.  In particular, we expect stability if
\begin{equation}
\max_i \|H(\vec{x}_i)\|^2\epsilon \in O( \delta),
\end{equation}
where $\delta\le Ae^{-\gamma N}$ is defined to be the error in the estimate of the unknown Hamiltonian and  $\gamma$ is approximately a constant function in $N$ for the test cases considered in the main body of the paper.  We therefore expect the algorithm to be stable if $\epsilon$ is chosen as above and hence, it will suffice to use a number of simulations in an update that approximately scales as
\begin{equation}
N_{\rm sim} \in \Theta\left(\frac{\max_i \|H(\vec{x}_i)\|^4|\{\vec{x}_i\}|}{\delta^2}\frac{\max_k \Pr(D|H(\vec{x}_k))(1-\Pr(D|H(\vec{x}_k)))}{\left( {\sum_k \Pr(D|H(\vec{x}_k))\Pr(H(\vec{x}_k))}\right)^2} \right).
\end{equation}
It is then easy to see from Markov's inequality that with high probability over the experiments the cost of any given update will be at most a constant multiple of the cost of the expectation value over all prior distributions $\Pr(\vec{x}_i)$ that appear in the learning process and all outcomes $D$ observed.  Therefore, our approximation to the total number of simulations required to learn the parameters within loss $\delta$ scales, with high probability, as
\begin{align}
N_{\rm total} &\in \Theta\left(\frac{N\max_i \|H(\vec{x}_i)\|^4|\{\vec{x}_i\}|}{\delta^2}\mathbb{E}\left(\frac{\max_k \Pr(D|H(\vec{x}_k))(1-\Pr(D|H(\vec{x}_k)))}{\left( {\sum_k \Pr(D|H(\vec{x}_k))\Pr(H(\vec{x}_k))}\right)^2}\right) \right)\nonumber\\
&\in \Theta\left(\frac{\log(1/\delta)\max_i \|H(\vec{x}_i)\|^4|\{\vec{x}_i\}|}{\gamma\delta^2}\mathbb{E}\left(\frac{\max_k \Pr(D|H(\vec{x}_k))(1-\Pr(D|H(\vec{x}_k)))}{\left( {\sum_k \Pr(D|H(\vec{x}_k))\Pr(H(\vec{x}_k))}\right)^2}\right) \right).
\end{align}

This suggests that QLE and IQLE may be efficient, given that $\gamma\in \Omega(\poly(1/n))$, $\max_j\|H_j\|\in O(\poly (n))$ and experiments that yield, with high probability, $\Pr(D|H(\vec{x}_k))\in O(1/\poly(n))$ are avoided.  We observed that these scalings are obeyed for the examples considered in the main body if IQLE and the particle guess heuristic are employed.  More complex examples may require local optimization of the guesses in order to avoid multi-modal prior distributions, which can be problematic for the PGH; however, we saw no benefit to local optimization for the Ising models considered previously.  Finally, the scaling predicted for $N_{\rm total}$ as a function of $\delta$ does not seem to be tight in the prior examples since $N_{\rm total}$ does not appear to strongly depend on $\delta$ in those examples.  A more careful analysis of the uncertainty is likely to reveal that our conditions for stability are unnecessarily pessimistic.
\end{document}